\documentclass[aip,apl,amsmath,amssymb,preprint,]{revtex4-1}
\usepackage{epsfig}
\usepackage{color}
\usepackage{graphicx}
\usepackage{dcolumn}
\usepackage{bm}

\begin{document}

\title{Forced generation of simple and double emulsions in all-aqueous system}

\author{Alban Sauret}
\affiliation{Department of Mechanical Engineering, University of Hong Kong, Pokfulam Road, Hong Kong} 
\affiliation{Institut de Recherche sur les Ph\'enom\`enes Hors \'Equilibre, UMR 7342, CNRS \& Aix-Marseille University,
49, rue F. Joliot-Curie, F-13013 Marseille, France}

\author{Ho Cheung Shum}\email[Electronic mail: ]{ashum@hku.hk}
\affiliation{Department of Mechanical Engineering, University of Hong Kong, Pokfulam Road, Hong Kong}

\date{13 April 2012}
\begin{abstract}
We report an easy-to-implement method that allows the direct generation of water-in-water (w/w) single emulsions. The method relies on direct perturbation of the pressure that drives the flow of the dispersed phase of the emulsions. The resultant inner jet is induced to break up into droplets due to the growth of the perturbation through Rayleigh-Plateau instability [L. Rayleigh, Proc. R. Soc. London {\textbf{29}}, 71Ð97 (1879)]; this leads to the formation of monodisperse droplets. By implementing this method on a modiÞed microfluidic device, we directly generate water-in-water-in-water (w/w/w) double emulsions with good control over the size and the number of encapsulated droplets. Our approach suggests a new route to apply droplet-based microfluidics to completely water-based systems.
\end{abstract}

\maketitle

\label{firstpage}


Recent advances in the generation of emulsion drops have led to applications in various fields such as food industry, cosmetics, drug delivery, and oil extraction.\cite{stone2004,squire2005,basaran2002,gunther2006,muschiolik2007} The ability to generate single and multiple emulsions with controlled morphology has been used for fabricating a variety of functional materials including microgels, liposomes, polymersomes, and colloidosomes.\cite{kim2007b,shah2008,shum2008,shum2011a,lee2009b}

However, the ability of producing monodisperse emulsions in some systems such as aqueous two-phase systems (ATPS), where the two aqueous phases have different properties, such as density, viscosity, and refractive index, is challenged by their low interfacial tensions.\cite{hardt2012} The tension for ATPS is typically less than $0.1$ mN$\,$m$^{-1}$,\cite{ziemecka2011} preventing the breakup of the jet.\cite{shum2010} An approach to induce droplet formation in these systems is to apply an external forcing to the flow at a given frequency. The resulting perturbation can induce the breakup of the jet into droplets through Rayleigh-Plateau instability.\cite{plateau1849,rayleigh1879,eggers2008} 
This has been accomplished through the use of an electric field in a flow-focusing microfluidic device,\cite{kim2007} piezoelectric actuator\cite{ziemecka2011} or multi-level rounded channels.\cite{lai2011} However, these methods are not always simple to implement on a microfluidic device, for instance, when electric fields cannot be used or when the device material does not allow straightforward incorporation of the functional components needed.\cite{utada2005}

In this Letter, we report a method to directly generate water-in-water (w/w) emulsions and water-in-water-in-water (w/w/w) double emulsions in glass microcapillary devices. To achieve this, we perturb the pressure that drives the flow of the dispersed phase in a controlled manner, disrupting the jet periodically. This results in monodisperse droplets whose size can be tuned. By implementing this approach on a modified microfluidic device, we directly generate w/w/w double emulsions with good control over their size and the number of encapsulated droplets.

\begin{center}
\begin{figure}[h!]
\begin{center}\includegraphics[width=11cm]{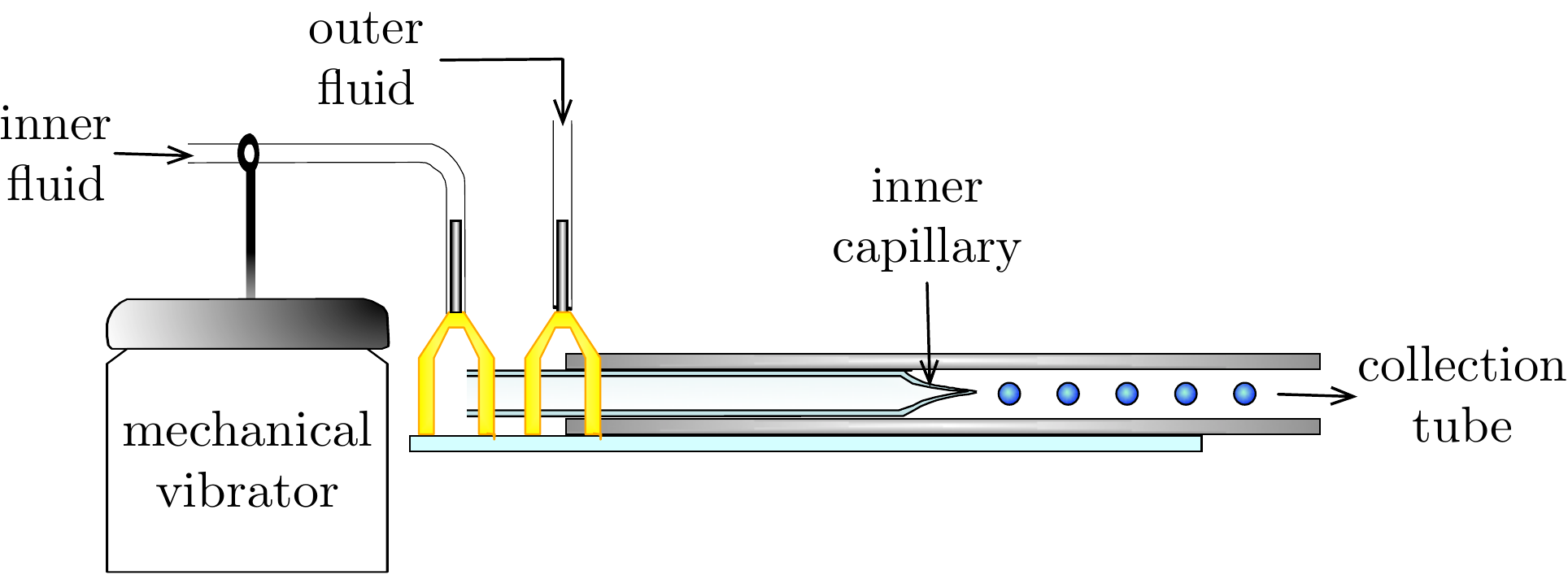}\end{center}
\caption{Schematic of the experimental setup. The flexible tubing directing the inner fluid into the capillary microfluidic device is connected to a mechanical vibrator that induces controlled pressure variation at the entrance of the inner capillary.}
\label{scheme_device}
\end{figure}
\end{center}

The experimental setup is made up of two coaxially aligned capillary tubes. A cylindrical inner capillary, with an approximate tip diameter of $30 \,\mu\text{m}$, is coaxially inserted in a square outer capillary, with an inner dimension of  $1 \,\text{mm}$. Two flexible tubings bring the inner and outer fluids in the capillaries at the flow rates $Q_{in}$ and $Q_{out}$ respectively. The inner phase is an aqueous solution of polyethylene glycol (PEG, MW=8000, $17\%$ wt) and the outer phase is an aqueous solution of dextran (T-500, MW=500 000, $15\%$ wt). The interfacial tension between the two phases is low, about $0.1$ mN$\,$m$^{-1}$.\cite{ziemecka2011} We connect a mechanical vibrator (PASCO Model SF-9324) to the tubing for injecting the inner fluid. The vibrator is controlled by an external generator for tuning the frequency in the range $[0.1;5000]\,\text{Hz}$ with a sinusoidal variation. Vibration of the tip of the inner capillary was not observed; the effect induced by the mechanical vibrator is only due to variation of pressure at the imposed frequency. This method enables precise control of the frequency of the pressure perturbation, which is key to the production of droplets with good control over droplet sizes. The experimental setup is shown in figure \ref{scheme_device}.

\begin{center}
\begin{figure}[h!]
\begin{center}\includegraphics[width=9cm]{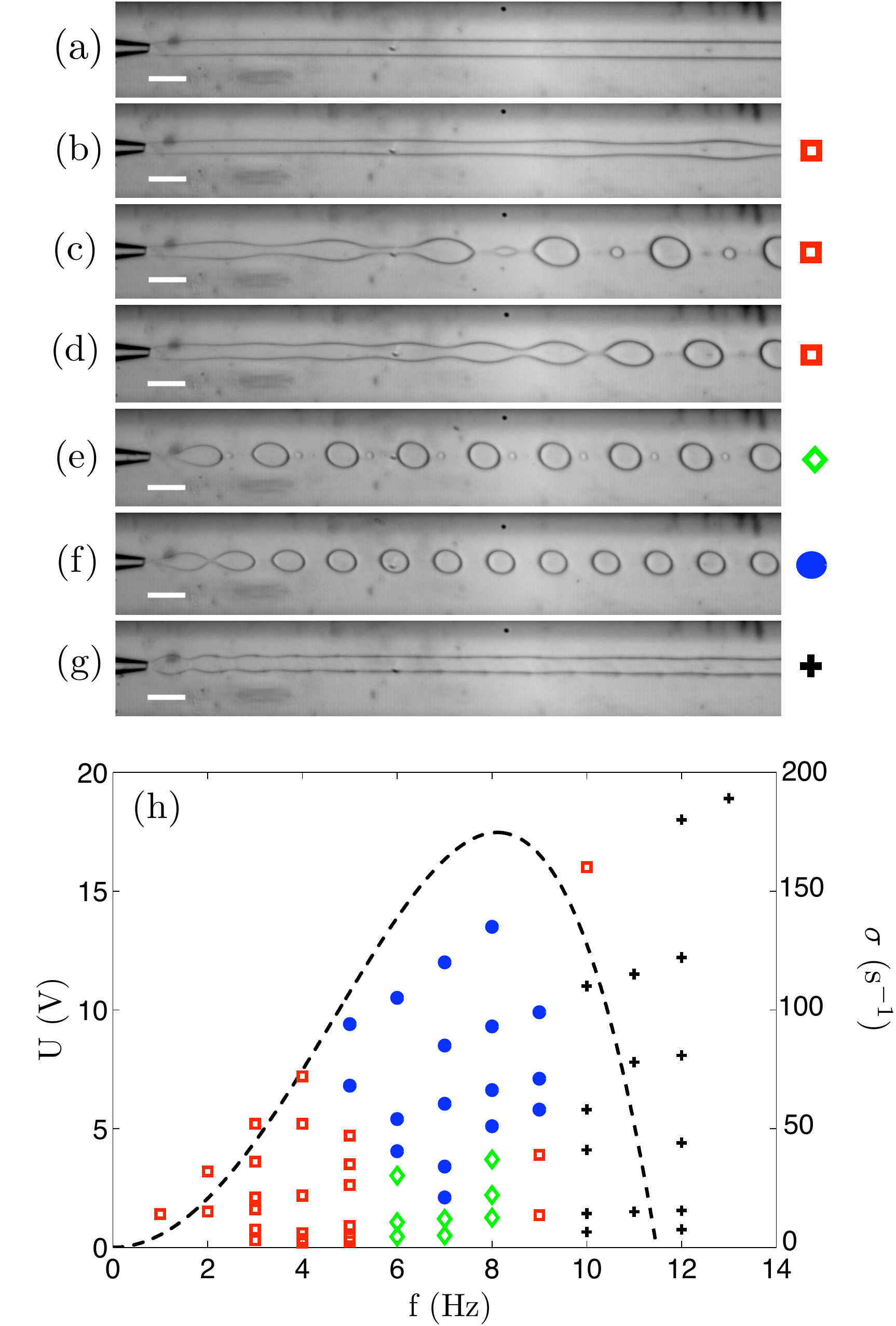}\end{center}
\caption{Different flow regimes as a function of the frequency of the forcing. The inner and outer fluid flow rates are respectively $Q_{in}=50\,\mu\text{L}\,\text{h}^{-1}$ and $Q_{out}=5000\,\mu\text{L}\,\text{h}^{-1}$. Optical image of the jets and drops observed (a) in the absence of external forcing, $f=0$ Hz, (b) at $f=3$ Hz, (c) $f=4$ Hz, (d) $f=6$ Hz, (e) $f=7$ Hz, (f) $f=8$ Hz, (e) $f=10$ Hz (scale bars are $200\,\mu$m). A state diagram of the different regimes as a function of the frequency of the forcing (in Hz) and the input voltage applied to the mechanical vibrator (in volts, left scale) is represented in (h). The dashed-dotted line is the growth rate of the Rayleigh-Plateau instability\cite{guillot2007} (right scale).}
\label{syst_f_epsil}
\end{figure}
\end{center}

To generate w/w emulsions, the dispersed phase is injected through a plastic flexible tubing connected to the mechanical vibrator, which oscillates at a specific frequency and induces pressure perturbation in the dispersed phase. The continuous phase flows through the square capillary. For a typical set of fluid flow rates of the dispersed phase and without perturbation, a jet exiting the inner capillary does not break up into droplets, as shown in figure \ref{syst_f_epsil}(a).\cite{shum2012} At the flow rates, $Q_{in}=50\,\mu\text{L}\,\text{h}^{-1}$ and $Q_{out}=5000\,\mu\text{L}\,\text{h}^{-1}$, a stable jet is visible and no droplet formation occurs. For a given input voltage, a low perturbation frequency leads to the formation of a wavy interface with a well defined wavelength between the two fluids, as shown in figure \ref{syst_f_epsil}(b) where a frequency of $f=3$ Hz is applied. The wavelength visible at the interface is directly linked to the frequency of excitation {for given inner and outer fluid flow rates and thickness of the inner jet $r_{jet}$ by the relation $\lambda=Q_{in}/(f\,\pi\,{r_{jet}}^2)$ and the velocity of the fluid of the inner jet is given by $v_{jet}=Q_{in}/(\pi\,{r_{jet}}^2)$. However}, the jet does not break up into droplets at this frequency except sporadically and very far from the nozzle. As the frequency of perturbation is increased to $f=4$ Hz, the jet starts to break up into droplets near the inner capillary nozzle with satellite droplets connecting the larger droplets (figure \ref{syst_f_epsil}(c)). At slightly higher frequency of $f=6$ Hz, the jet directly breaks up into droplets {with no visible satellites droplets} (figure \ref{syst_f_epsil}(d)). When the frequency reaches $f=7$ Hz, monodisperse droplets with smaller satellite droplets form by dripping at the capillary tip (figure \ref{syst_f_epsil}(e)). At an optimized frequency, f=8 Hz in this case, monodisperse droplets are induced in a dripping regime (figure \ref{syst_f_epsil}(f)). If the frequency of perturbation becomes larger than a critical value, which is  f=10 Hz in the present case, the growth rate becomes zero. Therefore, the perturbations visible in the part of the jet near the nozzle are quickly smoothed out as the jet flows downstream (figure \ref{syst_f_epsil}(g)). Based on our systematic investigation of the effects of perturbation frequency and input voltage on the flow regimes, monodisperse droplets formed directly at the tip of the inner capillary can be achieved only for a sufficiently high input voltage (typically more than $U=3$ V) and a small range of frequency ($f\in[5;9]$ Hz). Under these conditions, the growth rate of the Rayleigh-Plateau instability\cite{guillot2007} is maximized at  $f=8$ Hz (black dotted line in figure \ref{syst_f_epsil}(h)). At low applied perturbation frequency, $f <4$ Hz, the droplets are formed at some distance from the tip of the inner capillary. Due to the slow rate of growth of the perturbation, the time and thus the distance from the tip to the point at which the jet breaks up become very large. At high frequency $f\geq 11$ Hz, no droplets are produced. The inability to generate droplets is predicted by the Plateau criterion for the breakup of the jet, which states that when the wavelength $\lambda$ of the perturbation is smaller than the circumference of the jet, i.e., for $\lambda <2\,\pi\,r_{jet}$, no perturbation can grow and thus the jet does not break up. In terms of frequency of perturbation, it leads to the inequality $f>Q_{in}/(2\,\pi\,{r_{jet}}^2)\simeq 11$ Hz.\cite{eggers2008} The amplitude of pressure perturbation is tightly related to the amplitude of the shaking of the inner tubing, which is tuned by varying the input voltage of the vibrator. For a given input voltage $U=3$ V, droplets are formed directly at the tip of the inner capillary only for $f=7$ Hz. However, as the input voltage is increased up to $U=6$ V, the range of frequency for generation of droplets is increased to $f\in[5;9]$ Hz. However, the exact relationship between the amplitude of pressure perturbation and that of shaking in an oscillating flexible tube remains challenging.\cite{whittaker2010} A state diagram summarizing the morphologies at different perturbation frequencies and input voltages is presented in figure \ref{syst_f_epsil}(h).

\begin{center}
\begin{figure}[h!]
\begin{center}\includegraphics[width=9cm]{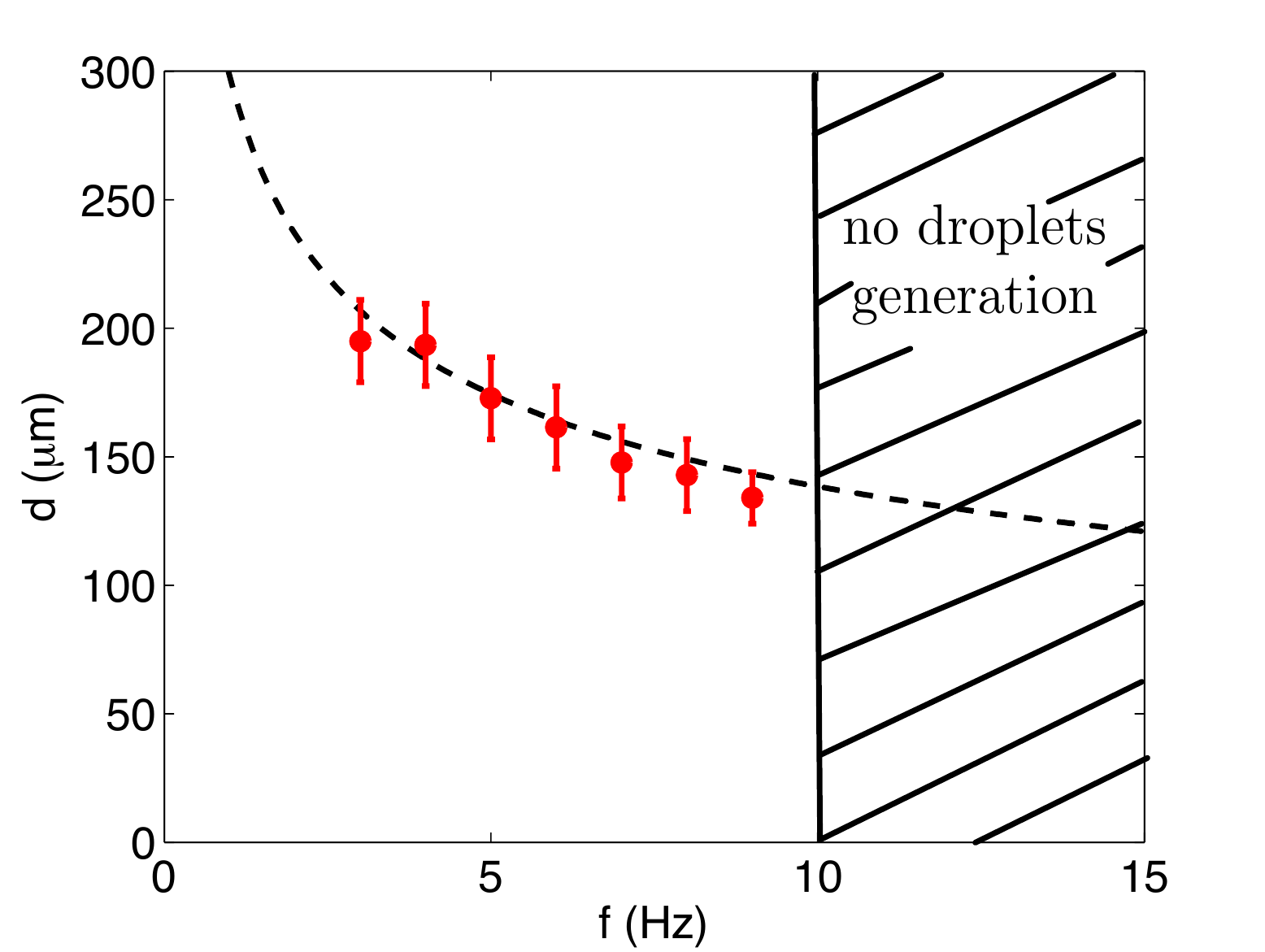}\end{center}
\caption{Diameter of the droplets as a function of the perturbation frequency for inner and outer fluid flow rates of $Q_{in}=50\,\mu\text{L}\,\text{h}^{-1}$ and $Q_{out}=5000\,\mu\text{L}\,\text{h}^{-1}$ respectively. The black dashed line is the scaling law. Experimental diameters of the generated drop are indicated by the red circles.}
\label{syst_d}
\end{figure}
\end{center}

The present experimental approach also enables control over the size of the emulsion drops. When induced by perturbation, the jet is forced to break up at a wavelength imposed by the frequency of applied perturbation. The volume of one droplet of diameter $d$ is calculated as $V_{drop}=\pi\,d^3/6$. As the frequency of the droplet formation matches the perturbation frequency $f$, the volume of the jet that contributes to one droplet is $V_{jet}=\pi\,{r_{jet}}^2\,v_{jet}/f$. The flow rate of the dispersed phase is defined as $Q_{in}=\pi\,{r_{jet}}^2\,v_{jet}$. Mass conservation implies that the volume $V_{drop}$ and $V_{jet}$ are equal. Therefore, the diameter of one droplet scales as $d=(6\,Q_{in}/f)^{1/3}$.\cite{utada2007} This expression is in excellent agreement with the experimental measurements shown in figure \ref{syst_d}.

\begin{center}
\begin{figure}[h!]
\begin{center}\includegraphics[width=11cm]{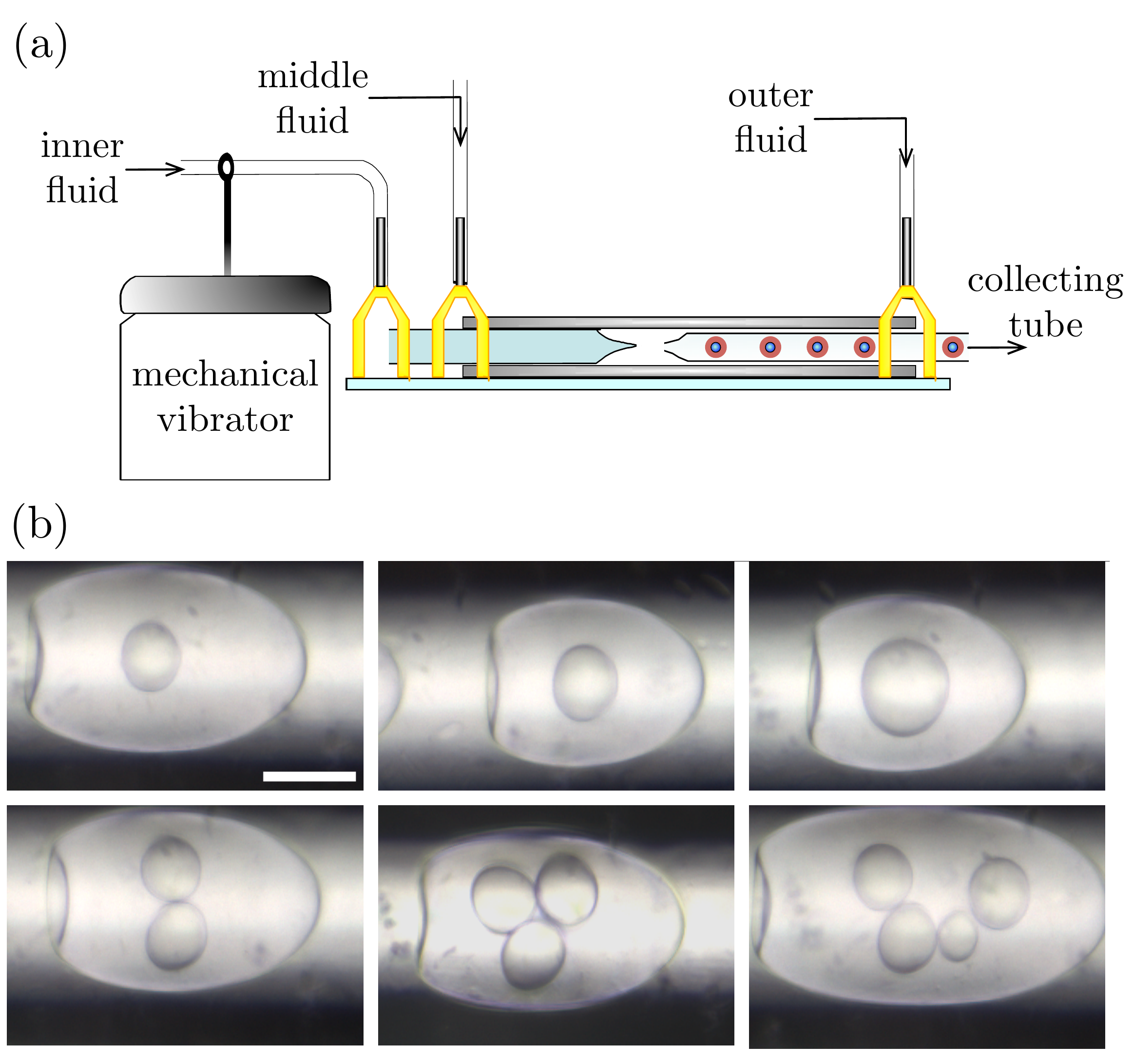}\end{center}
\caption{(a) Schematic of the setup for the generation of double emulsion. (b) w/w/w double emulsion, the size of the inner droplets as well as their numbers can be controlled by varying the frequency of excitation and the flow rates. Scale bar is $100\,\mu$m.}
\label{double_emulsion}
\end{figure}
\end{center}

A further challenge in materials engineering, apart from generation of monodisperse w/w emulsions is the preparation of w/w/w double emulsions. All-aqueous core-shell droplets have recently been generated through a process of spontaneous phase separation.\cite{ziemecka2011b} Unlike in classical droplet microfluidic approaches, where the species to be encapsulated are separated from the continuous phase by a middle shell phase, undirected emulsification processes likely lead to limited efficiency when applied to encapsulate cells or other active ingredients in a given phase. Our approach for fabricating controlled w/w emulsions can be adapted for the formation of w/w/w double emulsion using a modified glass microcapillary device\cite{utada2005} (see figure \ref{double_emulsion}(a)). In this case, aqueous PEG solution forms the middle phase while aqueous dextran solution forms both the inner and outer phases. The mechanical vibrator is connected to the inner plastic tubing as in the formation of simple w/w emulsion. The formation of the inner droplets within the middle phase helps modify the shape of the middle/outer interface, as shown previously in a multiphase flow with low interfacial tension.\cite{shum2010} Consequently, this facilitates the breakage of the middle jet into droplets. By tuning the outer fluid flow rate, we can match the breakup frequency of the middle jet to that of the inner droplets. Thus, droplets containing fixed number of inner droplets, that is, w/w/w double emulsion, are achieved. Tuning the frequency and the input voltage as well as the flow rates of both phases lead to a good control over the size of the inner droplets and the number of the inner droplets encapsulated in the middle fluid (see figure \ref{double_emulsion}(b)). 

In this Letter, we report a new approach that enables the generation of monodisperse all-aqueous single and double emulsions. This is achieved by inducing a variation of pressure of the dispersed phase. Our approach is easy to implement, low-cost and allows a good control over the flow morphologies. Moreover, we also demonstrate the first direct generation of w/w/w double emulsion. The resulting emulsions are not stabilized by any surface-active agents and destabilize through coalescence upon contact with neighboring droplets. These unconventional emulsions are not easily stabilized by conventional amphiphilic agents. While we illustrate the concept using capillary microfluidics, the understanding acquired is applicable to other microfluidic approaches, such as two-dimensional poly(dimethyl siloxane)-based  microfluidics. The ability to generate single and double emulsions in an all-aqueous environment in the absence of any organic solvent creates important opportunities to fabricate completely biocompatible materials with low environmental risks using droplet microfluidics. With this approach, the established methods for oil-water-based emulsion can be adapted to all-aqueous emulsions, which have great potential for biomedical, pharmaceuticals, food and cosmetics application that require high degree of biocompatibility.

\vspace{0.5cm}

We gratefully acknowledge financial support from the Seed Funding Programme for Basic Research from the University of Hong Kong (201101159009).


\end{document}